# Classical density functional theory for associating fluids in orienting external fields

Bennett D. Marshall[1], Walter G. Chapman[1] and Margarida M. Telo da Gama[2,3]

[1]*Department of Chemical and Biomolecular Engineering, Rice University, 6100 S. Main, Houston, Tx 77005*
[2]*Centro de Física Teórica e Computacional, Avenida Professor Gama Pinto 2, P-1649-003 Lisbon,Portugal*
[3]*Departamento de Física, Faculdade de Ciências da Universidade de Lisboa, P-1749-016 Lisbon, Portugal*

We develop a classical density functional theory (DFT) for two site associating fluids in spatially homogeneous external fields which exhibit orientational inhomogeneities. The Helmholtz free energy functional is obtained using Wertheim's thermodynamic perturbation theory and the orientational distribution function is obtained using DFT in the canonical ensemble. It is shown that an orientating field significantly enhances association by ordering the molecules thereby reducing the entropic penalty of association. It is also shown that association enhances the orientational order for fixed field strength.

The statistical mechanics of associating fluids has been the focus of much development for many decades. In the 1980's the search for primitive models of the hydrogen bond was the catalyst for advancement. Development of a theory for hydrogen bonding had to account for both the anisotropic nature of the pair potential as well as bond saturation. In a series of four classic papers[1-4] Wertheim developed a multi – density cluster expansion incorporating these basic features of the hydrogen bond. Since its development, Wertheim's theory has been widely applied in the modeling of homogeneous hydrogen bonding fluids[5-8] and, more recently, to study the self assembly and phase behavior of patchy colloid fluids[9, 10]. A similar multi – density approach for associating fluids with spherically symmetric attractions[11] has had application to highly asymmetric electrolyte solutions[12] as well as equations of state for dipolar fluids[13]

Patchy colloids are colloids with discrete attractive patches giving orientation dependant potentials with limited valence.[9] Recently, a number of synthetic methods[14-16] have been developed to produce patchy colloids. This gives researchers the ability to program the self assembly of these colloids into colloidal molecules[16], Kagome lattices[17] and network fluid phases[18], which consist of self- assembled chains with complex phase behavior including a re-entrant diagram[19]. The fluid structure, ranging from the molecular (angstroms) to the mesoscopic (microns) scale, can be tuned, for example, through the intermolecular interactions or the temperature[19] and the response to external fields is expected to be large (macroscopic) even for weak fields. Detailed calculations, however, are still lacking as the interactions with e.g. an orienting field require the account, within Wertheim's theory, of the orientational degrees of freedom of the monomers.

In addition to homogeneous systems, Wertheim's theory has found wide application in the field of inhomogeneous associating fluids in the form of classical density functional theory (DFT).[20] Each of these approaches has considered inhomogeneities in position only (confined fluids, interfaces etc…) and have neglected the possibility of inhomogeneities in orientation. Of course there are many cases in which associating fluids are placed in an external field which acts on orientation. First and foremost, due to anisotropy of the pair potential, any inhomogeneity in position is also an inhomogeneity in orientation. Other examples include associating dipolar molecules[21], patchy colloids in uniform electric fields and magnetic Janus particles in magnetic fields[22]. With the wide ranging success of Wertheim's theory in describing bulk associating fluids and associating fluids which exhibit spatial inhomogeneities, it seems timely to extend the approach to include the effects of orientating external fields. This will be the subject of this letter.

Specifically we will consider the case of spherical molecules of diameter $d$ with two short ranged directional association sites, labeled *A* and *B*, on opposite poles of the sphere (centers of the sites separated by $180°$). This gives an intermolecular potential which is the sum of a hard sphere and association contribution. We restrict attractions such that *A* attracts *B*, but there are no *AA* or *BB* attractions. For the association potential between site *A* and *B* we will employ conical square well association sites giving the potential[23-25]

$$\phi_{AB}(12) = \begin{cases} -\varepsilon_{AB}, & r_{12} \leq r_c \text{ and } \beta_{A1} \leq \beta_c \text{ and } \beta_{B2} \leq \beta_c \\ 0 & \text{otherwise} \end{cases} \quad (1)$$

where $1 = \{\vec{r}_1, \Omega_1\}$ represents the position $\vec{r}_1$ and orientation $\Omega_1$ of sphere 1, $\beta_{A1}$ is the angle between the site orientation vector of site *A* on sphere 1 and the vector connecting the center of the two spheres $\vec{r}_{12}$. The size of the association site is governed by the angle $\beta_c$ and the range of the interaction by the critical radius $r_c$. We will consider a spatially uniform external potential $\phi_E(1) = \phi_E(\Omega_1)$ such that density is a function of orientation only. We will approximate the free energy at the level of first order perturbation theory which assumes that association sites are singly bondable, steric hindrance between sites is negligible, no ring formation and finally that Wertheim's single chain approximation holds.[4] The restriction that each site is singly bondable is accurate for patch sizes $\beta_c < 30°$ (at $r_c = 1.1d$ )[26] and the effects steric hindrance and ring formation are small for large bond angles.[27] For this two site case the intrinsic Helmholtz free energy functional in TPT1[4] simplifies to

$$\frac{A - A_{HS}^{EX}}{Vk_BT} = \int \rho(\Omega)\left(\ln X_o(\Omega) + \ln(\rho(\Omega)\Lambda^3) - \frac{X_A(\Omega)}{2} - \frac{X_B(\Omega)}{2}\right)d\Omega \quad (2)$$

where $A_{HS}^{EX}$ is the hard sphere reference system *excess* Helmholtz free energy, $\rho(\Omega)$ is the density of spheres with orientation $\Omega$ and $X_o(\Omega) = X_A(\Omega)X_B(\Omega)$ is the fraction of molecules with orientation $\Omega$ which are not bonded at either site (monomer fraction). The term $\Lambda$ is the thermal De Broglie wavelength and finally $X_A(\Omega)$ is the fraction of molecules with orientation $\Omega$ *not* bonded at site *A* and is given by the mass action equation

$$\frac{1}{X_A(\Omega_1)} = 1 + \int d\Omega_2 d\vec{r}_{12} \rho(\Omega_2) X_B(\Omega_2) f_{AB}(\Omega_1, \Omega_2, r_{12}) g_{HS}(r_{12}) \qquad (3)$$

where $f_{AB}(12) = \exp(-\phi_{AB}(12)/k_B T) - 1$ is the association Mayer function and $g_{HS}(r_{12})$ is the pair correlation function of the hard sphere reference system. Note, in the hard sphere reference system the orientations of spheres are uncorrelated $g_{HS}(\Omega_1, \Omega_2, r_{12}) = g_{HS}(r_{12})$. To obtain the mass action equation for site *B* simply switch $A \rightarrow B$ in Eq. (3).

To obtain the density profile we employ classical density functional theory in the canonical ensemble. We construct the appropriate functional by adding the field contribution to the intrinsic free energy and adding a Lagrange multiplier term to enforce a constant number of spheres to obtain

$$\frac{A}{k_B T V} + \int d\Omega \rho(\Omega) \frac{\phi_E(\Omega)}{k_B T} - \lambda\left(\int \rho(\Omega) d\Omega - \rho_b\right) \qquad (4)$$

where $\rho_b$ is the bulk density, $\phi_E(\Omega)$ is the external field and $\lambda$ is the Lagrange multiplier. Minimizing Eq. (4) with respect to $\rho(\Omega)$ and enforcing the constraint $\rho_b = \int \rho(\Omega) d\Omega$ to eliminate $\lambda$ we obtain the orientational distribution function $\psi$

$$\psi(\Omega) = \frac{\rho(\Omega)}{\rho_b} = \frac{\frac{1}{X_o(\Omega)} \exp\left(-\frac{\phi_E(\Omega)}{k_B T}\right)}{\int \frac{1}{X_o(\Omega')} \exp\left(-\frac{\phi_E(\Omega')}{k_B T}\right) d\Omega'} \qquad (5)$$

In Eq. (5) the prefactor $1/X_o(\Omega)$ modifies the Boltzmann distribution favoring orientations which maximize association. The orientational average of any quantity $Q$ is then given by $\langle Q \rangle = \int \psi(\Omega) Q(\Omega) d\Omega$. Now we will restrict our attention to an external potential which has the form of a dipole in an electric field

$$\frac{\phi(\Omega)}{k_B T} = \frac{\phi(\cos\gamma)}{k_B T} = -\frac{\mu E}{k_B T}\cos\gamma = -C^* \cos\gamma \tag{6}$$

where $\gamma$ is the angle between the orientation vector and the field $\vec{E}$ (see Fig. 1) and $\mu$ is the dipole moment. Here we define the orientation vector of sphere $j$ as the vector which passes through the center of the $A$ patch $\vec{r}_A^{(j)}$ (hydrogen association site).

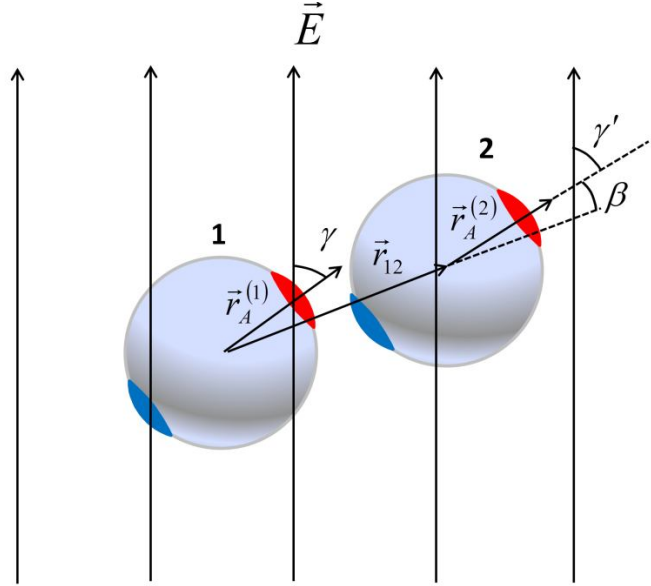

**Figure 1:** Diagram of associating spheres in an external field $\vec{E}$. The orientation of sphere $j$ is defined as the orientation of the center of patch $A$, denoted by the site orientation vector $\vec{r}_A^{(j)}$.

To complete the theory we must carry out the integration in Eq. (3). To perform this integral in a rigorous manner a numerical 4D integral (two spatial angles and two orientation angles) would need to be performed. A more computationally efficient method, although not exact, is to carry out the orientational integration $I$ for sphere 2 in an orientational reference frame centered on the site $A$ orientation vector $\vec{r}_A^{(1)}$ of sphere 1 ($\vec{r}_{12} // r_A^{(1)}$ in Fig. 1), and then assume that the total integral in Eq. (3) is given by this integral $I$ multiplied by the solid angle of the patch. Following this approach only a double integral must be evaluated, which is much more computationally convenient. Carrying out this process we obtain

$$\frac{1}{X_A(\cos\gamma)} = 1 + 2\pi(1-\cos\theta_c)\xi \hat{f}_{AB} \int_{\cos\theta_c}^{1} \int_{0}^{2\pi} d\cos\beta \, d\alpha \rho(\cos\gamma')X_B(\cos\gamma') \quad (7)$$

where $\hat{f}_{AB} = \exp(\varepsilon_{AB}/k_B T) - 1$ and $\xi = \int_{d}^{r_c} g_{HS}(r)r^2 dr$ which we approximate as[28] $\xi = (r_c - d)d^2 g_{HS}(d)$. We evaluate $g_{HS}(d)$ using the Carnahan and Starling result[29]. Finally, $\cos\gamma'$ is given by $\cos\gamma' = \cos\gamma\cos\beta - \sin\alpha\sin\beta\sin\gamma$ where $\beta$ and $\alpha$ are the polar and azimuthal angles $\vec{r}_A^{(2)}$ makes in a coordinate system whose $z$ axis is parallel to $\vec{r}_{12}$. Performing a similar integration for $X_B(\cos\gamma)$ we find $X_B(\cos\gamma) = X_A(\cos\gamma)$. This is to be expected since $X_A(\cos\gamma)$ is the fraction of molecules *with orientation* $\cos\gamma$ which are not bonded at site $A$.

To test the theory we perform *NVT* Monte Carlo simulations to measure the average orientation of spheres with respect to the field $\langle\cos\gamma\rangle$ and the average chain length of associated chains $\langle L_C \rangle = 1/\langle X_A \rangle$. Simulations were allowed to equilibrate for ~$10^9$ trial moves and averages were taken over an additional ~$10^9$ trial moves. We used $N = 864$ spheres for each simulation; however, additional simulations were performed using 2048 spheres to verify the absence of any system size effects. In this work we use the potential parameters $r_c = 1.1d$ and $\beta_c = 27°$ such that sites are only singly bondable.

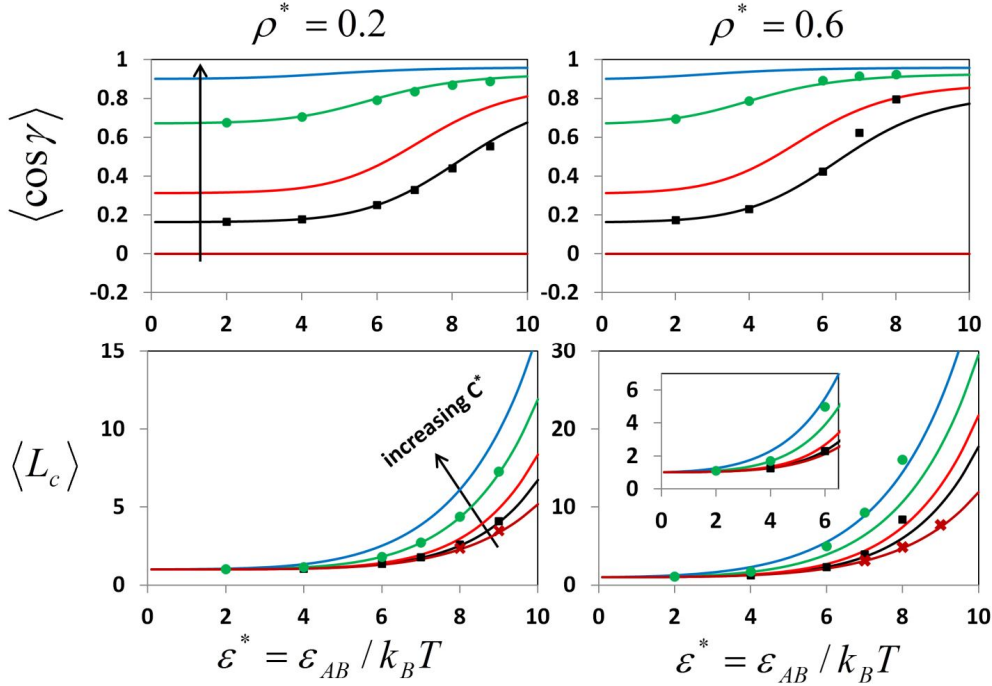

**Figure 2:** Average orientation with respect to the field $\langle \cos\gamma \rangle$ and average chain length $\langle L_C \rangle$ versus reduced association energy $\varepsilon^*$ for $\rho^* = 0.2$ (left) and $\rho^* = 0.6$ (right). Curves give theoretical predictions for, from bottom to top, $C^* = 0, 0.5, 1, 3$ and $10$. Symbols give simulation results for $C^* = 0$ (crosses), $C^* = 0.5$ (squares) and $C^* = 3$ (circles). Inset on bottom right panel shows results for $\langle L_C \rangle$ at $\rho^* = 0.6$ and $\varepsilon^* < 6.5$. Arrows point in direction of increasing $C^*$.

Figure 2 compares theoretical predictions and simulations for both $\langle \cos\gamma \rangle$ and $\langle L_C \rangle$ as a function of $\varepsilon^* = \varepsilon_{AB}/k_B T$ for various reduced field strengths $C^* = \mu E/k_B T$ at both low $\rho^* = \rho_b d^3 = 0.2$ and high $\rho^* = 0.6$ density cases. The general trend (excluding the zero field case) is that increasing the reduced association energy $\varepsilon^*$ increases $\langle \cos\gamma \rangle$ showing that association increases order in the system. For instance, for $\rho^* = 0.2$ and $C^* = 0.5$ increasing $\varepsilon^*$ from 0 to 10 results in a threefold increase in $\langle\cos\gamma\rangle$. We also note that increasing the field strength $C^*$ while holding $\varepsilon^*$ constant results in a significant increase in the average chain length $\langle L_C\rangle$. The increase in $\langle L_C\rangle$ with increasing $C^*$ is due to the fact that the field pre-orients the spheres thereby decreasing the penalty paid in decreased orientational entropy when a bond is formed. The relationship between $\langle\cos\gamma\rangle$ and $\varepsilon^*$ seems to be a synergistic effect between the decrease in energy obtained for both bond formation and orientational alignment with the field.

For the low density case $\rho^* = 0.2$, theory and simulation are in excellent agreement over the full range of $\varepsilon^*$, validating the approximation of the mass action equation (7). For the high density case $\rho^* = 0.6$, theory is in good agreement with simulation for $\varepsilon^* < 6$ and becomes less accurate for larger $\varepsilon^*$. The disagreement between simulation and theory for this case is likely due to the fact that in the simulations the oriented chains create a nematic type phase where long ranged order assists in the formation of chains by restricting associated chains to reptating in a quasi 1D tube formed by neighboring associated chains. The 1D nature of the tube assists in guiding chain ends together to form longer chains. This type of effect cannot be captured in the current theory due to the fact that Eq. (2) is obtained in Wertheim's single chain approximation which neglects interactions between associated clusters beyond that of the reference fluid.[30] This order can be observed in the simulation snapshots given in Fig. 3. As can be seen, the field orders and lengthens the associated clusters in the direction of the field.

Interestingly, for $\rho^* = 0.6$ the theory again becomes accurate for the quantity $\langle\cos\gamma\rangle$ at large $\varepsilon^*$ and $C^* = 3$. For the case $C^* = 0.5$ the nematic ordering effect, which is not captured in the theory, assist in orienting the spheres resulting in the underprediction of $\langle\cos\gamma\rangle$; however, for $C^* = 3$ the orientation of the spheres is completely dominated by the field. It is in this realm the theory again becomes accurate for $\langle \cos \gamma \rangle$ at large $\varepsilon^*$. However, due to the ordering of clusters, the error in $\langle L_C \rangle$ persist.

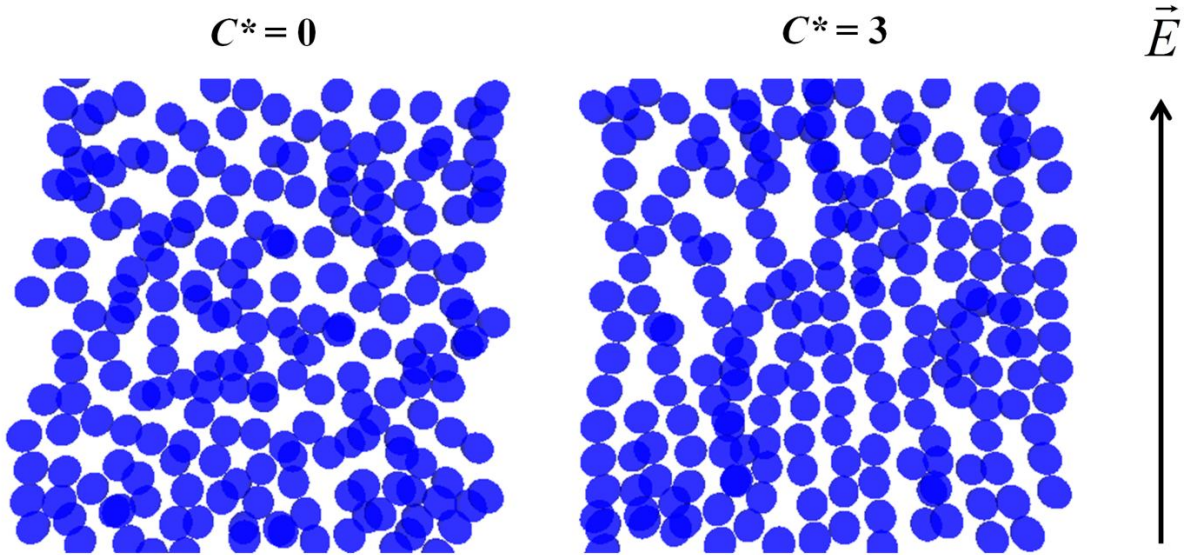

**Figure 3:** Simulation snapshots at a density of $\rho^* = 0.6$ and association energy of $\varepsilon^* = 8$ with no field $C^* = 0$ (left) and with field $C^* = 3$ (right). Snapshots are taken by displaying spheres located in a layer of $1.5d$ at the edge of the simulation cell. Due to periodic boundary conditions and neglect of spheres not in the layer, some clusters are longer than they appear.

We have developed a classical density functional theory for associating fluids in spatially homogeneous orientating fields. In this letter we have focused on the 2 site case where there are only *AB* attractions. If we included *AA* and *BB* attractions we should expect the same trends to hold, but association would increase in the fluid resulting in more order and longer chains. Also, extension to the general case[28] (in TPT1) of a set of association sites $\Gamma = \{A, B, C...\}$ is straight forward and leaves Eq. (5) unchanged with the monomer fraction now given by $X_o = \prod_{S \in \Gamma} X_S$. The broken orientational symmetry, as a result of the external field, allows for the enforcement of bond angle constraints in the theory, which is not possible in TPT1 otherwise.[27] This fact alone hints that the theory should be accurate for the case where there is more than two association sites. Also, we note that the axially symmetric case studied in the paper is probably the most difficult case to study theoretically due to the fact that, in this case, association induces long range order of the clusters. Possible extensions of this work include incorporation of the results of this letter with a bulk equation of state for dipolar molecular fluids[31] allowing for the study of the effect of electric fields on phase equilibria. Also, using the multi – density approach of Kalyuzhnyi and Stell[11], Kalyuzhnyi et al.[13] developed an equation of state for dipolar hard spheres. Combining their approach with the approach developed in this paper may allow for the development of a density functional theory for dipolar hard spheres in electric fields.

## Acknowledgments

The financial support of The Robert A. Welch Foundation Grant No. C – 1241 is gratefully acknowledged. Financial support from the Portuguese Foundation for Science and Technology (FCT) under Contracts nos. EXCL/FIS-NAN/0083/2012, PEst-OE/FIS/UI0618/2011, and PTDC/FIS/098254/2008 is also acknowleged.